\documentclass[prd,nofootinbib,preprint,superscriptaddress]{revtex4}

\usepackage{amsmath, amssymb, amsthm, graphicx, epsfig, fancyhdr,epsfig, slashed, mathrsfs}

\usepackage{tikzsymbols}
\usepackage{natbib}
\usepackage{float}

\usepackage{textcomp,bm,subfig,comment}

\usepackage{mathtools} 
\usepackage{hyperref}
\usepackage{tikz,xcolor}
\definecolor{lime}{HTML}{A6CE39}
\DeclareRobustCommand{\orcidicon}{
	\begin{tikzpicture}
	\draw[lime, fill=lime] (0,0) 
	circle [radius=0.2] 
	node[white] {{\fontfamily{qag}\selectfont \tiny ID}};
	\draw[white, fill=white] (-0.0625,0.095) 
	circle [radius=0.007];
	\end{tikzpicture}
	\hspace{-2mm}
}

\newcommand{\Msun}{M_\odot}

\foreach \x in {A, ..., Z}{\expandafter\xdef\csname orcid\x\endcsname{\noexpand\href{https://orcid.org/\csname orcidauthor\x\endcsname}
			{\noexpand\orcidicon}}
}

\newcommand{\be}{\begin{equation}}
\newcommand{\ee}{\end{equation}}
\newcommand{\bea}{\begin{eqnarray}}
\newcommand{\eea}{\end{eqnarray}}

\newcommand{\mc}{\mathcal}

\usepackage{array}
\newcolumntype{C}[1]{>{\centering\let\newline\\\arraybackslash\hspace{0pt}}m{#1}}

\def\lsim{\mathrel{\raise.3ex\hbox{$<$\kern-.75em\lower1ex\hbox{$\sim$}}}}
\def\gsim{\mathrel{\raise.3ex\hbox{$>$\kern-.75em\lower1ex\hbox{$\sim$}}}}

\newcommand{\Mpc}{{\rm Mpc}}

\newcommand{\td}{{\rm d}}
\newcommand{\mpl}{M_{\rm P}}

\newcommand{\epsh}{\epsilon_{H}}
\newcommand{\epsv}{\epsilon_{V}}
\newcommand{\etah}{\eta_{H}}
\newcommand{\etav}{\eta_{V}}

\newcommand{\calH}{\mathcal{H}}

\newcommand{\mr}[1]{\mathrm{#1}}

\definecolor{Red}{rgb}{1,0,0}
\definecolor{Blue}{rgb}{0,0,1}
\definecolor{Green}{rgb}{0,1,0}

\newcommand{\bb}{\begin{bf}} 
\newcommand{\eb}{\end{bf}}

\newcommand{\ben}{\begin{enumerate}} 
\newcommand{\een}{\end{enumerate}} 
\newcommand{\bi}{\begin{item}} 
\newcommand{\ei}{\end{item}} 
\newcommand{\bc}{\begin{center}} 
\newcommand{\ec}{\end{center}} 
\newcommand{\bl}{\begin{Large}} 
\newcommand{\el}{\end{Large}} 
\allowdisplaybreaks

\begin{document}
\title{Cosmological probes of Grand Unification: \\\it{Primordial Blackholes  \& scalar-induced Gravitational Waves}}
%
%\title{Probing the scale of Grand Unification via Primordial Blackholes as dark matter \& scalar-induced Gravitational Waves}
%
%\title{Probing the scale of Grand Unification via \\ Primordial Blackholes \& induced Gravitational Waves}
%
\author{Anish Ghoshal\orcidA{}}
\email{anish.ghoshal@fuw.edu.pl}
%\affiliation{INFN Rome, Italy}
\affiliation{Institute of Theoretical Physics, Faculty of Physics, University of Warsaw, \\ ul. Pasteura 5, 02-093 Warsaw, Poland}

\author{Ahmad Moursy\orcidB{}}
\email{a.moursy@fci-cu.edu.eg}
\affiliation{
Department of Basic Sciences, Faculty of Computers and Artificial Intelligence, Cairo University, Giza 12613, Egypt.
}
\author{Qaisar Shafi\orcidC{}}
\email{qshafi@udel.edu}
\affiliation{Bartol Research Institute, Department of Physics and Astronomy \\
University of Delaware, Newark, DE 19716, USA}
%
%\vspace{-10.9cm}
\begin{abstract}
\textit{We investigate the inflationary cosmology involving an SU(5) GUT (grand unified theory) singlet scalar with non-minimal coupling to the Ricci scalar. In this scenario the scale of grand unification is set by the inflaton vev when the inflaton rolls down its potential towards its minimum $v$, thereby relating inflationary dynamics to GUT symmetry breaking with a prediction $r \simeq 0.025$ for the tensor-to-scalar ratio to be tested by the next generation CMB experiments. We show in this inflationary framework involving inflection-point how a suitable choice of parameters in $SU(5)$ leads to a bump in the scalar power spectrum with production of Primordial Blackholes (PBH) of masses $10^{17}-10^{18}$g ($ 10 - 100 \Msun$). We derive the constraints on the self quartic and mixed quartic couplings of the inflaton in SU(5) that are consistent with the inflationary analysis. Moreover, we also show that this scenario leads to large amplitude induced second-order tensor perturbations propagating as Gravitational Waves (GW) with an amplitude $\Omega_{\rm GW}h^2 \sim 10^{-17}$ and peak frequency $f_{\rm peak} \sim$ (0.1 - 1) Hz, which can be detected in the next generation GW observatories like LISA, BBO, ET, etc. Thus, we unify the $SU(5)$ framework with PBH via inflection-point inflation showing how the upcoming measurements of PBH and GW will enable us to probe the scale of $SU(5)$ symmetry breaking, and thereby complementing the laboratory based experiments. We also discuss scenarios involving the Pati-Salam and Trinification gauge groups and its impact on quartic and mixed-quartic couplings that may lead to PBH and detectable GW signals.}
\end{abstract}
\maketitle
%
%%%%%%%%%%%%%%%%%%%%%%%%%%%%%%%%%%%%%%%%%%%%%%%%%%%%%%%%%%%%%%%%%
\section{Introduction}
\label{sec:intro}
%%%%%%%%%%%%%%%%%%%%%%%%%%%%%%%%%%%%%%%%%%%%%%%%%%%%%%%%%%%%%%%%%

One of the most striking predictions of Grand Unified Theories (GUTs) 
is proton decay 
\cite{Weinberg:1979sa,Wilczek:1979hc,Weinberg:1980bf,Weinberg:1981wj,Sakai:1981pk,Dimopoulos:1981dw,Ellis:1981tv}, and
 Super-Kamiokande has set stringent constraints on the typical decay channels such as  $p\to \pi^0 e^+$, $K^+ \bar{\nu}$ with the proton lifetime exceeding $10^{34}$ years~\cite{Abe:2014mwa,Miura:2016krn}. This approximately translates into a bound on the GUT symmetry breaking scale to be $M_{\rm GUT}$ to be $> 5 \times 10^{15}$ GeV. There are even more very exciting prospects ongoing and during the current decade, thanks to the upcoming large-scale and large volume neutrino detectors, namely the experiments like DUNE \cite{Acciarri:2015uup}, Hyper-Kamiokande \cite{Abe:2018uyc} and JUNO \cite{An:2015jdp}, which promise to improve the lower bound on $M_{\rm GUT}$ by an order of magnitude or so, or even more excitingly possibly detect proton decay. 

Besides the motivations of UV-completion (where all the SM gauge couplings are unified) GUTs also formed the basis of the first proposal for cosmic inflation, an accelerated expansion of the early universe, which essentially resolves the horizon and the flatness problems of big bang cosmology, as well as provide the initial seed of density fluctuations that grow into the inhomogeneous universe that we observe today \cite{Guth:1980zm,Sato:1980yn,Linde:1981mu,Petrosian:1982mt}\footnote{Later on, inflation was studied in the context of gravity effective theories like the Starobinsky scenario \cite{Nariai:1972zz,Starobinsky:1980te}.}. Such inflation driven by GUTs turned out to be unsuccessful due to measurements of the CMB, however the quantum generation of the primordial fluctuations seeding the large scale structure (LSS) of the Universe was a successful scenario. Whether or not the origin of inflationary cosmology be of particle physics, the rapidly increasing data from cosmological precision measurements, particle physics experiments and astrophysical observations lead us to the quest to find a coherent picture of the early Universe based on particle physics to begin with. Inflationary studies based on conformal GUT theories employed Coleman-Weinberg potential for a GUT singlet scalar inflaton field with minimal coupling to gravity \cite{Shafi:1983bd,Lazarides:1984pq}. It was soon followed by studies of GUT models containing topologically stable cosmic strings \cite{Kibble:1982ae,Shafi:1984tt} and intermediate scale monopoles \cite{Lazarides:1980cc,Senoguz:2015lba,Lazarides:2019xai,Maji:2022jzu, Lazarides:2022ezc}, and in GUT models such as $SO(10)$ these may survive an inflationary epoch. Later on, GUT models involving dark sector physics were also considered \cite{SravanKumar:2018tgk,Biondini:2020xcj,Kawai:2015ryj}.

Going beyond the minimal coupling to gravity, the Standard Model Higgs inflation and scalar fields with non-minimal coupling to gravity (the Ricci scalar) \cite{Bezrukov:2007ep} provide naturally flat inflaton potentials to scalar fields in general \cite{Khoze:2013uia,Kannike:2014mia,Rinaldi:2014gha,Salvio:2014soa,Kannike:2015apa,Kannike:2015fom,Barrie:2016rnv,Tambalo:2016eqr} that may drive inflation with predictions of spectral indices that are compatible with the current CMB measurements. In some models there appear certain characteristic features in the potential which may lead to production of huge scalar fluctuations that may seed formation of primordial black holes (PBHs) \cite{Novikov:1979} when length scales around the critical point re-enter the Hubble horizon after inflation has ended. Moreover, such PBHs may even constitute part or all of the dark matter (DM) in the universe \cite{Chapline:1975ojl, Dolgov:1992pu, Jedamzik:1996mr, Ivanov:1994pa, GarciaBellido:1996qt, Yokoyama:1995ex, Ivanov:1997ia, Blais:2002nd,Heurtier:2022rhf, Choudhury:2013woa, Chen:2022usd,Kawai:2021bye,Kawai:2021edk}, thus providing a novel DM candidate \footnote{SM Higgs Inflation may also lead to detectable scalar-induced GW signals, see Refs.\cite{Drees:2019xpp}}. Moreover the recent detection of gravitational waves from black hole mergers by LIGO and Virgo \cite{TheLIGOScientific:2016pea} has ushered in a new age in GW astronomy of the study of PBHs \cite{Carr:1974nx, Carr:1975qj}, with several proposed models of inflation involving scalar fields \cite{Garcia-Bellido:2017mdw, Ezquiaga:2017fvi,Kannike:2017bxn, Germani:2017bcs,Motohashi:2017kbs,Gong:2017qlj, Ballesteros:2017fsr,Rasanen:2018fom} that could produce PBH dark matter in a mass range observable by current or in the upcoming gravitational wave experiments\footnote{There is an ongoing debate regarding the formation of PBH as DM with discrepancies coming from one-loop effects in cosmological perturbation theory, see Refs. \cite{Kristiano:2022maq,Riotto:2023hoz,Kristiano:2023scm,Riotto:2023gpm,Choudhury:2023hvf,Choudhury:2023rks,Choudhury:2023vuj,Choudhury:2023jlt,Franciolini:2023lgy}. In this paper we do not go into details of such issues and instead discuss if GUT models may yield PBH (dark matter or some fraction thereof), and provide detectable GW signals as a novel probe of such unification.}. These predictions are, of course, subject to the stringent constraints such as lensing and gamma-ray bursts and other astrophysical measurements at our disposal \cite{Carr:2009jm, Carr:2017jsz,Carr:2016drx}. Only few regions remain in the spectrum of PBH mass that still allow a sizeable PBH population. These are basically two narrow mass windows at $10^{18}$ g and $4\times10^{19}$ g, and a PBH mass window around $10^{34}\dots10^{35} \, \mr{g} \approx 25\dots100 \, \mr{M}_\odot$ which is very near to the LIGO/Virgo range that may allow for PBH to be DM. All initial PBH masses $<10^6$ g are however disallowed as they correspond to PBHs that evaporate before big bang nucleosynthesis (BBN) down to Planck scale relics without spoiling the baryon asymmetry \cite{Carr:2009jm, Carr:2017jsz}.

In addition to having PBH formation, second order tensor perturbations may be generated from the enhancement of the scalar curvature perturbations. This has been investigated in recent works, which lead to detectable GW signals in the upcoming GW detectors
\cite{Gong:2017qlj,Bartolo:2018evs,Hajkarim:2019nbx,Liu:2020oqe,Xu:2019bdp,Fu:2019vqc,Bartolo:2018rku,Clesse:2018ogk,Kohri:2018awv,Braglia:2020eai,Bhaumik:2022pil,Bhaumik:2022zdd,Chen:2023lou,Chatterjee:2017hru,Ferrante:2023bgz}. 

With the scale of GUT physics largely out of reach of laboratory experimental facilities, in this paper, we propose $SU(5)$ inflationary scenario which lead to large scalar perturbations and formation of PBH as the entirety of DM in the universe, and show how the GUT scale can be probed in the induced GW predictions with the upcoming GW experiments. The scenario is consistent with the lower bound on $M_{\rm GUT}$ experiments, proton decay, as well as theoretical constraints on the GUT parameter space arising from quantum corrections and the demand of unification of the SM gauge couplings.  We show that the scale of grand unification can be probed by cosmological observables including spectral distortions, PBH and GW predictions, and in future may lead to novel constraints that may pin down the GUT scale. Our discussion also extends to other unified models based on gauge groups $SU(4)_C \times SU(2)_L \times SU(2)_R $ \cite{Pati:1973uk} and $SU(3)_C \times SU(3)_L \times SU(3)_R $ \cite{Glashow:1984gc, Babu:1985gi,Dvali:1996fc}, in which case $M_{\rm GUT}$ can be lower than the standard scale of $5\times 10^{15}-10^{16}$ GeV.

\textit{The paper is organized as follows:} in Section II, we discuss the model of inflation, and in Section III we discuss the GUT symmetry breaking and the relation with inflationary parameters. In Section IV, we describe the scalar perturbations and the solution to the Mukhanov-Sasaki Equations resulting in the spike in the power spectrum. This leads to induced GWs  and the production of PBHs as DM which we discuss in Section V. Finally, we end with some discussion and outlook in Section VI.

\medskip

%%%%%%%%%%%%%%%%%%%%%%%%%%%%%%%%%%%%%%%%%%%%%%%%%%%%%%%%%%%%%%%%%

\section{An Inflation model with inflection point}
\label{sec:model}

%%%%%%%%%%%%%%%%%%%%%%%%%%%%%%%%%%%%%%%%%%%%%%%%%%%%%%%%%%%%%%%%%%
 
We consider an inflaton $\phi$ which is a GUT singlet and non-minimally coupled to gravity. The action in Jordan frame is given by
\bea
S_J= \int d^4x \sqrt{-g} \left[ \dfrac{1}{2} f(\phi) R- \dfrac{1}{2} g^{\mu\nu} \partial_{\mu}\phi \partial_\nu \phi -V_J(\phi)\right],
\eea 
with $V_J(\phi)$ being the scalar potential that has the most general form of a renormalizable potential, with non-zero vev of $\phi$, 
\bea
V_J(\phi) &=& \frac{1}{2} m^2\,(\phi - v)^2 - \frac{1}{3}\alpha\,\mu\,(\phi - v)^3 + \frac{1}{4}\lambda\,(\phi - v)^4 ,
\eea
and the non-minimal coupling to gravity is given by \cite{Bezrukov:2007ep}:
\be
f (\phi) = 1 + \xi \phi ^2 .
\ee
Here $\alpha, \lambda$ and $\xi$ are dimensionless couplings, while $m,\mu$ are dimensionful mass scales. The global minimum is located at $\langle \phi \rangle = v$, at which the potential is zero. 
We work in Planck units, where the reduced Planck mass $M_{\rm Pl}= 2.43 \times 10^{18}$ GeV is set to unity. In the Einstein frame $g^{\mu\nu}_E= f(\phi) g^{\mu\nu}$, the action has the form 
 \bea
S_E= \int d^4x \sqrt{-g_E} \left[ \frac{1}{2} R_E - \dfrac{1}{2} g_E^{\mu\nu} \partial_{\mu}\sigma \partial_\nu \sigma -V_E(\sigma(\phi))\right],
\eea 
where  the  field $ \sigma $, with canonical kinetic terms, is defined in terms of $\phi$ via 
\bea
\left(\frac{d\sigma}{d\phi} \right)^2 &=& \frac{1}{f(\phi)}+ \frac{3}{2} \left( \frac{f'(\phi)}{f(\phi)} \right)^2 \nonumber\\
&=& \frac{1+\xi (1+6\xi ) \phi^2 }{(1+\xi \phi^2)^2}, 
\eea
and the inflationary potential $V_E$ is given in terms of $\phi$ by
\bea
V_E(\phi) = \frac{V_J(\phi)}{f(\phi)^2}
= \frac{\frac{1}{2} m^2\,(\phi - v)^2 - \frac{1}{3}\alpha\,\mu\,(\phi - v)^3 + \frac{1}{4}\lambda\,(\phi - v)^4}{\Big(1+\xi\, \phi ^2\Big)^{2}} .
\label{eq:pot}
\eea
\begin{figure}[t!]
    \centering
    \includegraphics[width=0.8\linewidth]{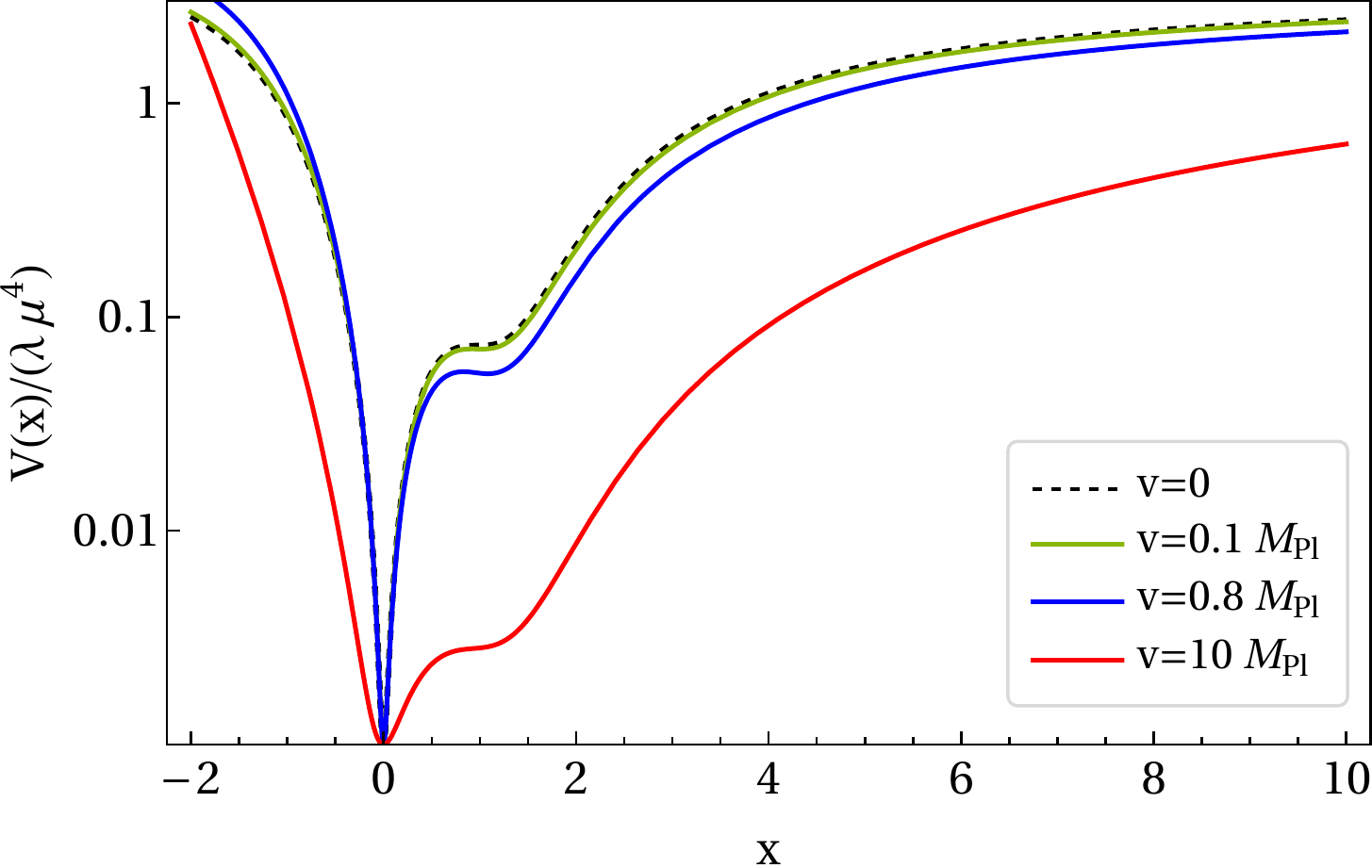}
    \caption{\label{fig:potx}\it Plot of the inflation potential~(\ref{eq:potx}) on $y$-axis versus the inflaton values $x$. Different colors correspond to different values of the vev $v$.}
\end{figure}
with the re-definitions, $x=(\phi - v)/\mu$, $m^2=\lambda\,\mu^2$, $a=\alpha/\lambda$ and $b=\xi\,\mu^2$, the potential \label{{eq:pot}} can be recast in the form
\be
V(x) = \frac{\lambda\,v^4}{12}\ \frac{x^2(6 - 4\,a\,x + 3\,x^2)}{ \left[ 1 + b \,\left(\dfrac{v}{\mu } +x \right)^2\right]^2}\,,
\label{eq:potx}
\ee
 where we have dropped the subscript $E$, for simplicity. This potential features an asymptotically flat direction for very large $x$, which is suitable for driving inflation. For $x\gg 1$, $V \to V_0 = \dfrac{\lambda v^4}{4 b}$.
 %%%%%%%%%%%%%%%%%%%%%%%%%%%%%%%%%%%%%%%%%%%
\subsection{Inflection points and ultra-slow roll regime}\label{sec;inflection}
 We are interested in inflection points of the potential~\ref{eq:potx}, $V''(x)=0$, around which the inflaton encounters ultra-slow roll (USR) regime, and therefore we require that $V'(x)\approx 0$. We analyze the critical values of the potential~\ref{eq:potx} by solving the equation $V'(x)=0$. One solution to the latter equation represents the true minimum at $x=0$, and the others are solutions to the cubic equation 
 \be
c_0 + c_1 x + c_2\,x^2+c_3\,x^3=0\ .
\label{eq:thirdorder}
\ee
with
\bea
c_0=1 \,, \hspace{0.5cm} c_1= -a, \hspace{0.5cm} 
c_2= \frac{3 \mu ^2 + b \left(3 v^2-2 a \mu  v-3 \mu ^2\right)}{3 \left(b v^2+\mu ^2\right)} , \hspace{0.5cm}
c_3= \frac{b \mu  (a \mu +3 v)}{3 \left(b v^2+\mu ^2\right)}
\eea 
\begin{figure}[H]
    \centering
    \includegraphics[width=0.7\linewidth]{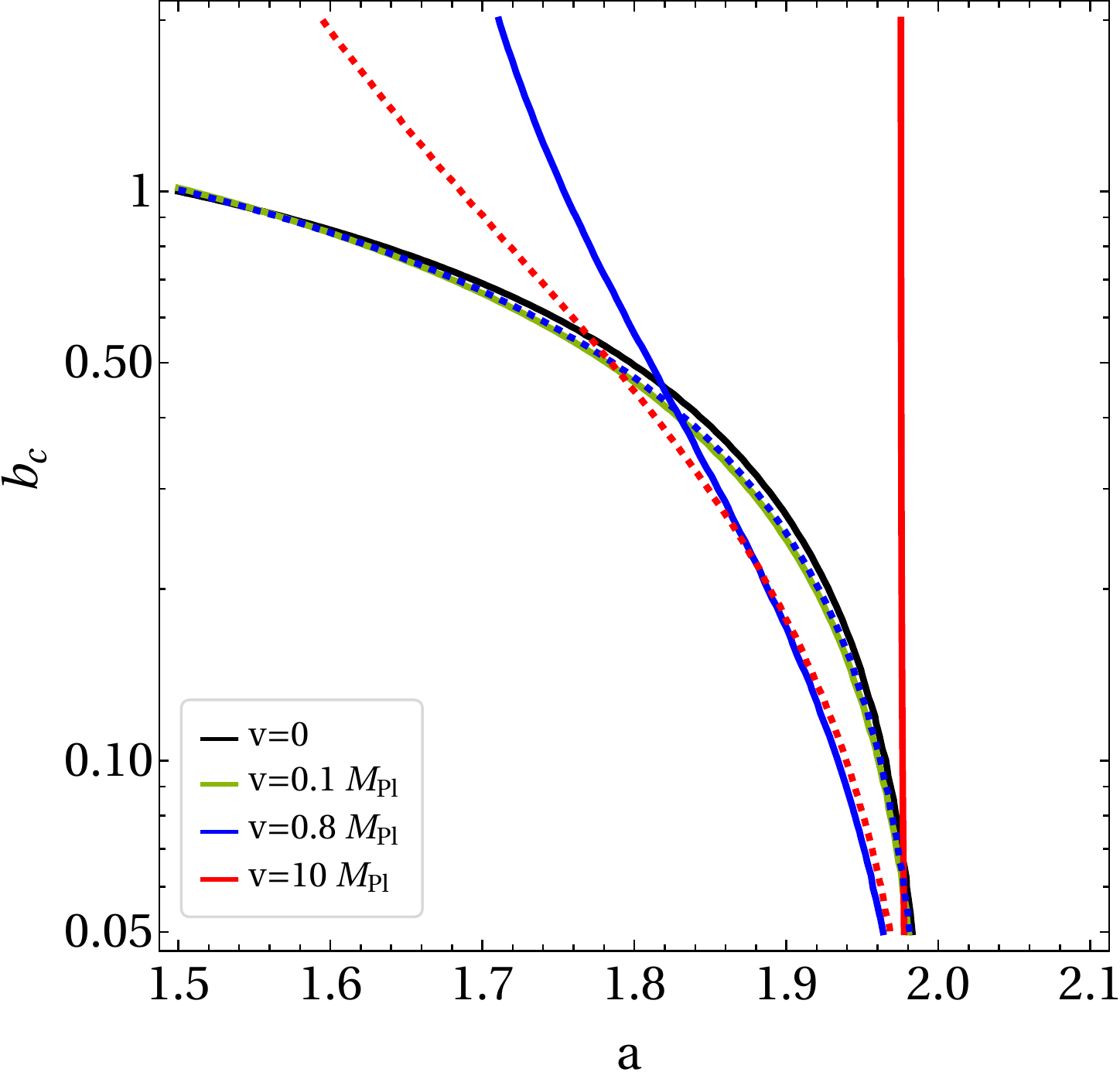}
    \caption{\label{fig:bc-a}\it The change of critical value $b_c$ versus $a$, with  different values of $v$ represented by different colors. The solid curves correspond to $\mu^2=0.5$, dashed curves correspond to $\mu^2=0.01$, and dotted curves correspond to $\mu^2=5$.}
    
\end{figure}
Generally, there are three solutions, $x_1$ and $x_{2,\,3} = x_0 \pm i\,y_0$, with 
\begin{align}
x_1  & =  -\frac{c_2}{3 c_3} -\frac{1}{3 c_3}\left( \Theta  (c_i) + \frac{D}{\Theta  (c_i)}\right),\\
x_0  & =    -\frac{c_2}{3 c_3} +\frac{1}{6 c_3}\left( \Theta  (c_i) + \frac{D}{\Theta  (c_i)}\right),\\
y_0  & =   \frac{1}{2 \sqrt{3} c_3}\left( \Theta  (c_i) - \frac{D}{\Theta  (c_i)}\right) .
\end{align}
Here
\bea
\Theta(c_i)^3= 
\left(   C - \sqrt{C^2-D^3} 
\right),
\eea 
with $D= c_2^2-3 c_1 c_3 = c_2^2+3 a c_3 $ and $C=\dfrac{1}{2}\left(2 c_2^3-9 c_1 c_3 c_2+27 c_0 c_3^2 \right)$. Clearly, $x_1<0$, and $x_2,x_3$ are real if,  $\Theta = C^{1/3} = \sqrt{D}$. In this case $x_2 =x_3=x_0$ will be a simultaneous solution for $V''(x)=0$ as well, namely, 
\be
x_0=\dfrac{-c_2 + D^{1/2}}{3c_3}.
\ee
For vev,  $v=0$, $c_0=1 \,, c_1= -a, 
c_2= 1-b\,,
c_3= \dfrac{ab}{3}$, and therefore our case reduces to the model described in Ref. \cite{Blanco-Pillado:2013qja}.
% In the Limit when $v\ll \mu$, we can expand $c_2,c_3$ in powers of $\dfrac{v}{\mu}$, 
%\bea
%c_2 = (1-b)-\frac{2 a b}{3}  \dfrac{v}{\mu}+ \cdots \,, \hspace{1cm}
%c_3 = \frac{a b}{3}+ b \, \dfrac{v}%{\mu}+ \cdots .
%\eea
%

The potential~\ref{eq:potx} has five independent parameters, namely, $(\lambda,\, \mu, \, a, \, b, \, v)$. However,  the existence of an inflection point depends only on $(a,\,b , \, \mu,  \, v)$. In particular, the desired inflection point exists at a critical value for $b=b_c(a,v)$, and hence the inflaton vev $v$ plays an important role in the dynamics around the inflection point. This is one of the main differences from \cite{Blanco-Pillado:2013qja} in which $v=0$ and accordingly $b_c$ is independent of $v,\mu$
\be
b_c(a)   =  1 - \frac{1}{3}\,a^2 + \Delta(a) \,, \hspace{1cm}
\Delta(a)  =  \frac{a^2}{3}\,\left(\frac{9}{2a^2}-1\right)^{2/3} \,.
\ee
This case is illustrated by the black curve in Fig.~\ref{fig:bc-a}. The other curves depicts the change in $b_c$ versus $a$ for different values of non-zero $v$ and different values of $\mu$.

The non-zero vev of $\phi$ will affect the slope and the curvature of  the potential during inflation and around the inflection point, as illustrated in Fig.~\ref{fig:potx}, where we have shown the effect of changing the the vev
$v$ on the shape of the potential. 
%\ag{Discussion on initial conditions..}
 
%%%%%%%%%%%%%%%%%%%%%%%%%%%%%%%%%%%%%%%%%%%%%%%%
\subsection{Slow-roll approximation and dynamics around the inflection point}\label{ssec:SRA}

In the slow-roll approximation (SRA), the slow-roll parameter $\epsilon_V$ and $\eta_V$ are given in terms of the potential as
\begin{align}\label{eq:eps}
\epsilon_{V} & = \frac{1}{2\mu^2}\left(\frac{V'(x)}{V(x)}\right)^2
\left(\frac{d\sigma}{d\phi} \right)^{-2} \\
\eta_{V} & = \frac{1}{\mu^2}\left(\frac{V''(x)}{V(x)}\right)
\left(\frac{d\sigma}{d\phi} \right)^{-2}
- \frac{1}{\mu}\frac{V'(x)}{V(x)} 
\left(\frac{d\sigma}{d\phi} \right)^{-3} \frac{d^2\sigma}{d\phi^2} 
\,,
\end{align} \label{eq:Ne}
and the number of e-folds is given by
\begin{align}\nonumber
N_{e} = \int_{x_e}^{x_*} \frac{1}{\sqrt{2\epsilon_{V}}} \,  \frac{d\sigma}{d\phi}\, dx
 \,.
\end{align}
\looseness=-1 

As indicated in Ref. \cite{Blanco-Pillado:2013qja}, as the inflaton passes through the inflection point ($x\to x_0$, $y_0 \to 0$), the integrand $dN/dx$ in eq.~(\ref{eq:Ne}) diverges in the SRA. Therefore, we consider a \emph{near-inflection point}, where $ b=b_c(a,\mu,v)-\varepsilon$ with the resonance parameter $0<\varepsilon\ll1$. Accordingly, one can control the number of e-folds spent at $x=x_0$ by choosing appropriate values of $\varepsilon,\mu$ and $ v$. Hence, a significant peak in the power spectrum can be produced.

%\medskip

\section{Gauge symmetry breaking  }
\label{sec:pt-infl}
%
%{\color{purple}
The part of the potential that dictates the interaction of the inflaton $\phi$ with some gauge symmetry breaking scalar $\chi_D$ is given by
\begin{align}\label{eq:inter-pot}
V_J(\phi,\chi_D) = - \frac{1}{2}\beta_D^2\phi^2\chi_D^2 + \frac{\lambda_D}{4}\chi_D^4, 
\end{align}
where we consider, for simplicity, $\chi_D$ as a canonically normalized real scalar field in $D$-dimensional representation of the gauge group. 
This potential induces a vacuum expectation value (VEV) for $\chi_D$ given by
\begin{align}\label{eq:vev-final}
\left<\chi_D\right> = (\beta_D/\sqrt{\lambda_D}) v \equiv M_{\rm SB},
\end{align}
when the inflaton reaches its true  minimum at $v$, therefore, 
\begin{align}
\beta_D  =   \dfrac{ \sqrt{\lambda_D} \, M_{\rm SB}}{v}.
\end{align}
\begin{figure}[H]
    \centering
    \includegraphics[width=0.7\linewidth]{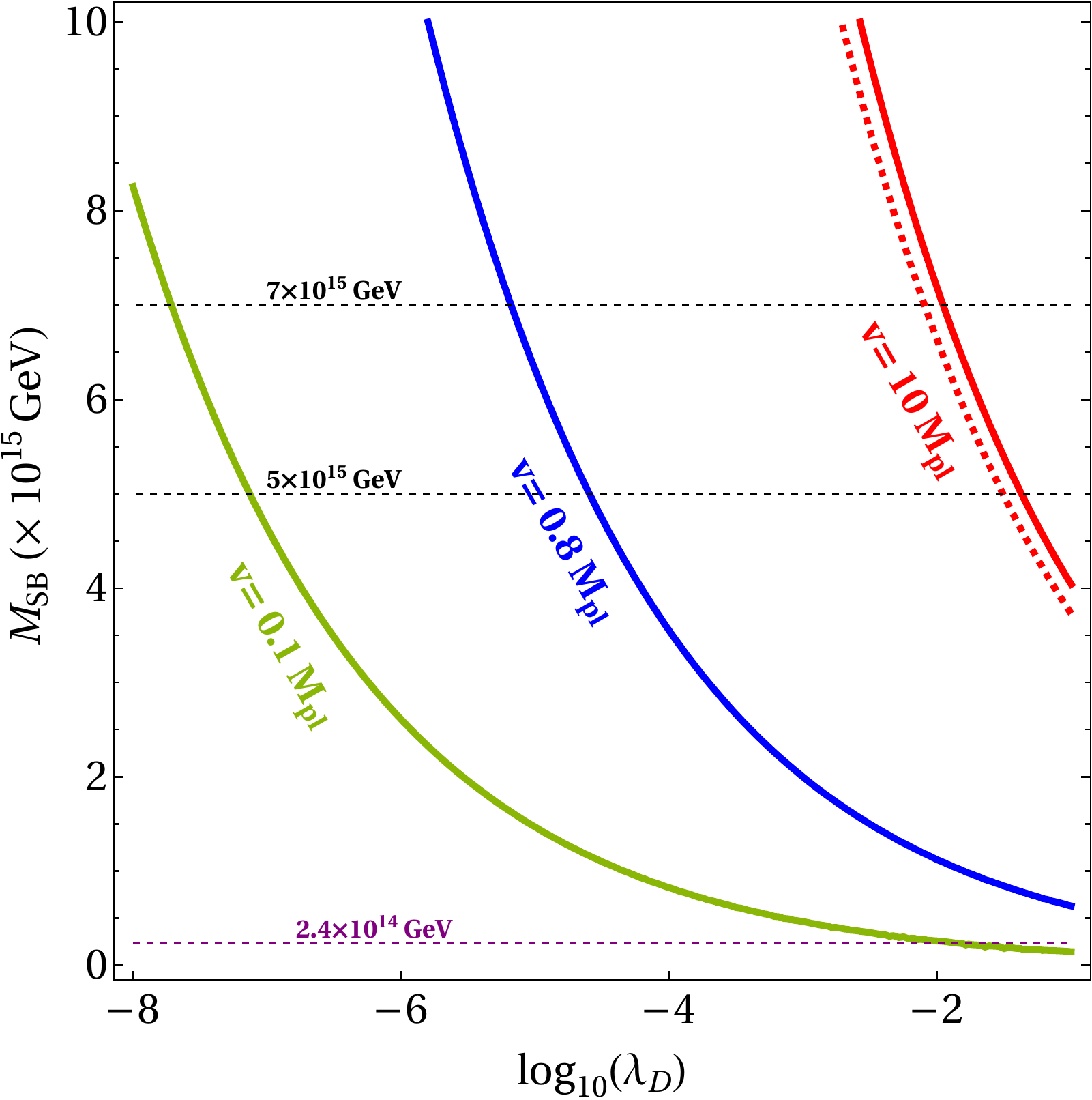}
    \caption{\label{fig:MSB}\it Allowed regions (below the curves) in the $M_{SB}-\lambda_D$ plane,  for different values of the inflaton vev. The solid (dashed) red curves correspond to the same vev $v= 10 \, {\text M_{Pl}}$, but have different values of tensor to scalar ratio 
    %$r \sim 0.0758 \, (0.064)$ 
    as shown in detail in Tables~\ref{tab:BPs},~\ref{tab:CMB-obs}.}
    
\end{figure}

In a Grand Unified (GUT) gauge group, such as  $SU(5)$ gauge symmetry, the gauge coupling is order unity, with the structure constant given by $\alpha_{GUT}=\dfrac{g_{GUT}^{2}}{4\pi} \sim \dfrac{1}{50}$, with a typical GUT scale $M_{\rm GUT} \sim  5\times 10^{15} - 1 \times 10^{16}$ GeV. Accordingly, the contribution to the  renormalization of the  quartic coupling $\lambda_D$ introduces a lower bound on the $\chi_D$ quartic coupling. For example, if we assume $\lambda_D \gtrsim 10^{-2}$, then this is translated to a lower bound on $\beta_D$, from Eq. \ref{eq:vev-final}
\be
    \sqrt{ 4\pi}    > \beta_D      \gtrsim       \, \dfrac{0.1 M_{\rm SB} }{v} ,
\ee
where the upper bound has been considered from the perturbativity conditions. Accordingly, a lower bound on the inflaton vev is given by $v\gtrsim 2.05 \times 10^{-4} \,\, M_{\rm Pl}$.

In this regard, let's discuss the role of the role of SU(5) Higgs $\chi_D$, during inflation, where the inflaton energy density during inflation is $\rho_\phi \sim V(\phi_*) \sim  10^{-9} \,\, M_{\rm Pl}^4$. The inflation scale is given by
 \be
V^{1/4}=10^{16} \left( \dfrac{r}{0.01} \right)\,\, {\rm GeV} .
 \ee
 According to the latest PLANCK and BICEP measurements on the upper bound on the tensor-to-scalar ratio (r), Planck 18 at 68\% CL:  0.1277 (at 2 $\sigma$) and at 95\% CL:  0.0684 (at 1 $\sigma$). While BICEP gives at 68\% CL: 0.0371 (at 2 $\sigma$)
and at 95\% CL: 0.0257 (at 1 $\sigma$). We depict the predictions of  of the \textit{tensor-to-scalar ratio} values $r $, for the benchmark points (BPs) chosen in Table~\ref{tab:BPs} \cite{Planck:2018vyg,BICEPKeck:2022mhb}. Likewise the energy scale of inflation rages between $2.5\times10^{16} -6.4\times10^{16}$ GeV for our choices of the benchmark points. On the other hand, the value of the SU(5) symmetry breaking scalar, during inflation, is $\chi_{D*} = \dfrac{\beta_D}{\sqrt{\lambda_D}} \, \phi_*$. Therefore, during inflation, 
 \bea
\rho_\chi &\sim & \frac{2 \, \beta_D^4 }{\lambda_D} \, \phi_*^4 =  2\lambda_D \, \left( \dfrac{  M_{\rm SB}}{v} \right)^4 \phi_*^4 
%&\approx&  4.3 \times 10^{-5} \, \lambda_D,
 \eea
 %
 %where we have used the parameters of Table~\ref{tab:BPs},  $v = 0.1 \, M_{\rm Pl}$,  $\phi_* = 16.55 \, M_{\rm Pl}$. Therefore, in case $\lambda_D\sim 10^{-5}$, 
 In this paper we are interested in the regions of parameter space where  inflation is driven by $\phi$ only. In this case, energy density of $\phi$ will be dominant during inflation if $\rho_\chi < \rho_\phi$.  
 In Fig.~\ref{fig:MSB}, we show the allowed  regions  under the curves in the $M_{\rm SB}-\lambda_D$ plane, for which the energy density of $\phi$ dominates the universe during inflation. The curves correspond to different values of inflaton vev. The  horizontal dashed black lines give different symmetry breaking scales $5\times10^{15}$ GeV and $7\times10^{15}$ GeV, that are typical GUT scales where $SU(5)$ is broken. At these values, $\lambda_D \gtrsim 0.01 $ for $v= 10 \, M_{\rm Pl}$ which is consistent with PLANCK-2018 measurements, while a tuning is required to $\lambda_D \gtrsim 6 \times 10^{-5} - 2.5 \times 10^{-5} $ for $v= 0.8 \, M_{\rm Pl}$, and $\lambda_D \gtrsim 2 \times 10^{-8} - 6.7 \times 10^{-8} $ for $v= 0.1 \, M_{\rm Pl}$ in order to be consistent with BICEP measurements. As a matter of fact, the quartic coupling $\lambda_D$ can be tuned $\lambda_D \sim 10^{-5} $ with an extended matter sector due to cancellations between gauge and Yukawa quantum corrections, if one considered an appropriate number of vector-like fermions that couple to $\chi_D$ \cite{Bhattacherjee:2017cxh}. Moreoever we have been focusing on the gauge group to be SU(5) that of GUT symmetry breaking scenario however the horizontal dashed purple line corresponds to symmetry breaking scale $2.4 \times10^{14}$ GeV that can be the scale of Pati-Salam unified theory with gauge symmetry ($SU(4)\times SU(2)_L\times SU(2)_R$)\footnote{We also note that some of the intermediate symmetry breaking chains involving SO(10) GUTs may give rise to cosmic strings which give rise to GW spectrum and the signals from induced GW spectrum presented in this paper and those may add up to give a combined signal while some breaking chains will not have generation of cosmic strings. All these depend upon the specific chain and the scale of symmetry breaking. Such a study is beyond the scope of present study. For details see \cite{King:2020hyd,King:2021gmj}.} or even trinification scenarios, with $\lambda_D \gtrsim 0.01$. The values of the symmetry breaking parameters as well as the inflationary observables are given in Table~\ref{tab:BPs} and~\ref{tab:CMB-obs}. However detailed analysis of such studies involving other gauge groups is beyond the scope of the present manuscript and we leave it for future publication.

\medskip

%%%%%%%%%%%%%%%%%%%%%%%%%%%%%%%%%%%%%%%%%%%%%%%%%%%%%%%%%%%%%%
\section{Inflationary perturbations and CMB}
\label{sec:theory}
%%%%%%%%%%%%%%%%%%%%%%%%%%%%%%%%%%%%%%%%%%%%%%%%%%%%%%%%%%%%%%
In this section, we study the inflationary perturbations and their cosmological consequences. In particular, we aim to investigate
the  generation of sufficiently large scalar fluctuations that may lead to PBH production within the realm of our inflationary scenario.  In this case, inflation proceeds in two or more phases of slow roll (SR) which should be separated in between  by a brief exit from the SR phase and a transient ultra slow roll (USR) phase. 
In this scenario, scalar modes that exit the horizon are enhanced  ~\cite{Leach:2000yw,Leach:2001zf}. The latter enhancement controls the position and height of the generated peak in the scalar power spectrum, and in turn may form PBHs with  varying mass scales and scalar induced GWs at different frequencies.

\subsection{Curvature Perturbations}

We start by giving the form of perturbed metric as follows
\begin{equation}
ds^2=a (\tau) ^2 \left[ -(1+2 \Phi)d \tau^2+\left( \left(1-2 \Psi\right)\delta_{ij}+\frac{1}{2}h_{ij}\right) dx^i dx^j \right]
\label{eq.metr}
\end{equation}
where $h_{ij}$ are the  tensor perturbations, while $\Phi$ and $\Psi$ are called the Bardeen potentials, which are equal in the conformal-Newtonian gauge.

Now we study the evolution of the scalar curvature perturbation  $\mc{R}_c$ by defining the Mukhanov field $v \equiv z \mc{R}_c$, with $ z \equiv a \dot{\phi}/H 
$, and $a$ is the scale factor \cite{Drees:2019xpp}.

 In this regard, the Fourier modes of $v$ evolve according to the MS equation~\cite{Mukhanov:1985rz,Sasaki:1986hm,Mukhanov:1990me},
\be\label{eq:MS}
%	\frac{d^2 u_k}{d \tau^2}+\left(  k^2 -\frac{1}{z} \frac{d^2 z}{d \tau^2} \right) u_k=0\,.
	v_{k}'' + \left(  k^2 -\frac{z''}{z} \right) v_{k}=0\,,
\ee
where the prime in the last equation means differentiation with respect to conformal time that is defined by, $\td \tau \equiv \td t / a$.\footnote{This shouldn't be confused with the prime in $V'$, which corresponds to a derivative with respect to the scalar field.} 
 The factor $ \left(  k^2 -\dfrac{z''}{z} \right) $ is considered to be an effective frequency $\omega_k^2(\tau)$. 
 It is convenient to express the factor $\dfrac{z''}{z}$ that is important to study the evolution of the modes, in terms of the SR parameters as follows: 
\bea\label{eq:zpp}
	\frac{z''}{z}
	&= \calH^2\left[ 2-\left(3-\epsh\right) \etav+\epsh \left(5+2\epsh-4\etah\right) \right] \, ,
\eea
where $\calH \equiv a'/a= aH$. It is worth mentioning that this expression is exact to all orders in the SR parameters. Also, we note that during the SR phase, the SR parameters are very small,  and therefore $z''/z \approx 2\calH^2$.

We recognize two phases for the evolution of the modes: the sub-horizon evolution phase where $k^2 \gg z''/z$, and the super-horizon evolution phase $k^2 \ll z''/z$.
In the sub-horizon limit, the mode $v_k$ behaves as a free field in flat spacetime. Therefore, in the Bunch-Davies vacuum, the normalized solution  goes like% in the limit $\tau \to - \infty$,  

\be\label{eq:BD_cond}
	v_{k}= \frac{1}{\sqrt{2 k}}e^{-ik \tau}\,
\ee
We use this free solution as the initial condition for the MS equation. 
On the other hand, in the super-horizon limit,  the general solution can be written as a linear combination \cite{Karam:2022nym}
\be 
v_k = A_{1,k} v^{(1)}_0 + A_{2,k} v^{(2)}_0
\ee 
where the mode $v^{(1)}_0$ grows during SR, while the mode $v_{0}^{(2)}$ decays during SR. They are expressed as \cite{Karam:2022nym}
\be\label{eq:SH_modes}
	v^{(1)}_0 \propto z, \qquad
	v^{(2)}_0 \propto z \int^{0}_{\tau} \frac{\td \tau'}{z(\tau')^2}.
\ee
Therefore $v^{(1)}_0$ will finally dominate and hence, the amplitude of the curvature perturbations is frozen \cite{Karam:2022nym}
\be 
|\mc{R}_{c\, k}| =\left\lvert v_k/z\right\rvert ={\rm const}.
\ee 

Now let's discuss what happens when SR is momentarily violated. In that case, the growing and decreasing $u_k$ modes can mix if $z$ decreases as advocated in \cite{Karam:2022nym}. Accordingly, the second mode may come to dominate momentarily, which may lead to a super-horizon evolution of $\mc{R}_c$. At the end of this temporary phase, $z$ starts to grow, and again,  $\mc{R}_c$ will be frozen. In this case, we can link the observable perturbations at horizon re-entry to the perturbations produced during inflation.

The primordial power spectrum is computed after the time of horizon crossing as studied in details in Ref. \cite{Drees:2019xpp,Karam:2022nym}, and is given by 
\be\label{eq:PR_from_u}
	{\cal P}_\xi
	=  \left. \frac{k^3}{2 \pi^2}\frac{|v_{k}|^2}{z^2} \right|_{\calH \gg k} 
	\stackrel{\rm SR}{\approx} \left. \frac{H^2}{8\pi^2 \mpl^2 \epsh} \right|_{k=\calH} \, .
\ee
Now, we can replace the Hubble SR parameters $\epsilon_H$ and $\eta_H$, by potential SR parameters $\epsilon_V$ and $\eta_V$. Therefore, the scalar spectral index $n_s$ and the tensor-to-scalar ratio $r$ are given at the leading order in SR  by 
\be \label{eq:SR_ns_and_r}
    n_s %\approx 1 + 2\etah - 4\epsh 
    \approx 1 + 2\etav - 6\epsv \, , \qquad 
    r %\approx 16\epsh 
    \approx 16\epsv \, .
\ee
The recent CMB observational constraints on inflationary observables are given at the pivot scale $k_{\rm pivot} = 0.05\ \Mpc^{-1}$~\cite{Planck:2018jri,BICEP:2021xfz}, as follows
\bea \label{eq:CMB_observables}
    A_s &= (2.10 \pm 0.03)\times 10^{-9} , \quad
    n_s = 0.9649 \pm 0.0042, \quad
    r < 0.036 \, .\qquad 
\eea
As a matter of fact, the curvature power spectrum is constrained at scales relevant for the CMB. Therefore, we define 
\be A_s \equiv {\cal P}_\xi(k_{\rm pivot}) = 2.1\times 10^{-9} \, .
\ee

\subsection{Inflationary background dynamics and spectrum of curvature perturbations}
Now we discuss the solution of the MS equation in order to calculate the primordial power spectrum. We first solve the background equation of motion of the  canonical inflaton field together with Friedmann equations,
\bea \label{eq:teqm}
\Ddot{\sigma} + 3H \dot{\sigma} + \dfrac{dV}{d\sigma} = 0, \nonumber\\
3 H^2 = \dfrac{1}{M_{\rm Pl}^2} \left [ \dfrac{1}{2} \dot{\sigma}^2 + V(\sigma)\right],
\eea 
where dot denotes the derivative with respect to the cosmic time. We define the number of efolds $N$ as $dN= H dt$ and rewrite equations~\ref{eq:teqm} to be merged into one equation  as follows:
\bea\label{eq:Neqm}
\dfrac{d^2 \sigma}{dN^2}+ 3 \dfrac{d\sigma}{dN} - \left[ 3 -\dfrac{1}{2} \left( \dfrac{d\sigma}{dN} \right)^2 \right] \dfrac{V'(\sigma)}{V(\sigma)} = 0\,,
\eea 
where we set $M_{\rm Pl}^2=1$. The Hubble SR parameters will be given in terms of $N$ as well, by
\bea 
\epsilon_H &\equiv&  \dfrac{1}{2} \dfrac{\dot{\sigma}^2}{H^2} =
\dfrac{1}{2} \left( \dfrac{d\sigma}{dN}\right)^2\,,
\\ \eta_H &\equiv& -\dfrac{\Ddot{\sigma}}{H \dot{\sigma}}  = \epsilon_H -\dfrac{1}{2 \epsilon_H} \dfrac{d \,\epsilon_H }{dN} \,.
\eea 
The scalar spectral index and the tensor to scalar ratio are then given respectively by 
\bea
n_s\approx 1-4 \epsilon_H + 2 \eta_H \,, \,\, r\approx 16 \epsilon_H \,.
\eea 
 In our numerical treatment, we rewrite the background equation of motion in terms of $x$ as follows\footnote{which is similar to Eq. (37) in \cite{Drees:2019xpp}}
\bea \label{eq:Neqmx}
&& v\left[ \dfrac{d\sigma}{d\phi } \dfrac{d^2 x}{dN^2} + 
v \dfrac{d^2\sigma}{d\phi^2 } \left( \dfrac{d x}{dN}\right)^2   \right] +
v  \dfrac{d\sigma}{d\phi } \dfrac{d x}{dN}  \left[3- \dfrac{1}{2} v^2  \left(  \dfrac{d\sigma}{d\phi } \right)^2  \left( \dfrac{d x}{dN}\right)^2  \right] +\nonumber \\
&& \dfrac{1}{v} \dfrac{d\phi}{d\sigma} \left[3- \dfrac{1}{2} v^2  \left(  \dfrac{d\sigma}{d\phi } \right)^2  \left( \dfrac{d x}{dN}\right)^2  \right] 
\dfrac{V'(x)}{V(x)}=0 .
\eea 
Moreover, we study the evolution of the curvature
perturbations by numerically solving the MS equation that is rewritten in terms of $N$ as \cite{Ballesteros:2017fsr}
\begin{equation} \label{eq:remuk}
\begin{split}
\frac {d^2v_k} {dN^2} &+ ( 1 -\epsilon_H) \frac {dv_k} {dN} \\
&+ \left[ \frac {k^2} {e^{2N}H^2} + ( 1 + \epsilon_H - \eta_H )
( \eta_H - 2 ) - \frac {d(\epsilon_H-\eta_H)} {dN} \right] v_k = 0 \,.
\end{split}
\end{equation}
We  can then compute the power spectrum at the end of inflation as 
\begin{equation} \label{msnend}
\mathcal{P}_\zeta(k) 
= \frac {k^3} {2\pi^2} \Big| \frac {v_k} {z} \Big|^2_{N=N_{\rm end}}\,.
\end{equation}
As we advocated above, we choose suitable initial conditions by assuming the Bunch--Davies vacuum at very early times \cite{Bunch:1978yq, Drees:2019xpp}:
\begin{equation} \label{initialconds}
\lim\limits_{\tau \to -\infty} v_k = \frac {{\rm e}^{-ik\tau}} {\sqrt{2k}}\,.
\end{equation}
Therefore, we have 
\begin{equation} \label{reini}
\mathrm{Re}(v_k) \Big|_{N=N_i} = \frac {1} {\sqrt{2k}} \, ;  \
\mathrm{Im}(v_k) \Big|_{N=N_i} = 0 \, ;
\end{equation}
\begin{equation} \label{imini}
\mathrm{Re} \left( \frac {dv_k} {dN} \right) \Big|_{N=N_i} =0 \, ; \
\mathrm{Im} \left( \frac {dv_k} {dN} \right) \Big|_{N=N_i} =
- \frac {\sqrt{k}} {\sqrt{2}a(N_i) H(N_i)}\,.
\end{equation}

where $N_i$ is the initial value of $N$ where we start the numerical integration of the MS equation.

%

%%%%%%%%%%%%%%%%%%%%%%%%%%%%%%%%%%%%%%%%%%
\subsection{Numerical Results}
%%%%%%%%%%%%%%%%%%%%%%%%%%%%%%%%%%%%%
 %
 \begin{table}[H]
 \begin{center}
\begin{tabular}{c | c | c | c | c | c | c | c| c}
 \hline\hline
    &    $v({\text M_{\rm Pl} })$   & $\mu^2 ({\text M_{\rm Pl}^2} )$ & $a$ & $b_c$ & $\varepsilon$  &  $\lambda$  & $\lambda_D$ & $ M_{SB}$(GeV)\\
\hline
%BP1 & $10^{-3}$& $0.5$  & $1.7824423$  & $0.5311$  &  $8.94 \times 10^{-4}$ &  $4.35 \times 10^{-9}$ \\
%\hline
 BP1 &   $0.1$  & $ 0.5$ &  $1.83457037$ & $0.3895$  & $3.24 \times 10^{-2}$  &  $2.7 \times 10^{-9}$  & $10^{-4}$ & $10^{15}$ \\
 \hline
BP2 &   $0.8$  & $ 2$ &  $1.9090835$ & $0.198991$  & $ 10^{-7}$  &  $6.1 \times 10^{-11}$  & $10^{-3}$ & $2\times 10^{15}$ \\
\hline
BP3 &   $10$  & $ 5$ &  $1.97677052$ & $0.06562$  & $ 10^{-6}$  &  \begin{tabular}{c}  $ 5.5 \times 10^{-12}$  \\\  $ 4.1 \times 10^{-12}$ \end{tabular}  & $10^{-2}$ & $6.8\times 10^{15}$ \\
 \hline
 \hline
\end{tabular}
\end{center}
\caption{\label{tab:BPs} \it Three benchmark points for the model parameters used to generate the distributions presented in Figs.~\ref{fig:ps1},~\ref{fig:SIGW1} and~\ref{fig:PBH1}. } 
\end{table}
 \begin{table}[h!]
 \begin{center}
\begin{tabular}{c | c | c | c | c | c }
 \hline\hline
    &  $n_s$  & $r$   & $A_s$ & $N_{\rm CMB}$ & $\phi_*$ \\
\hline
%Ob(BP1) & $0.946$  &  $0.0259$  & $2.1795 \times 10^{-9}$  &  $77.14$  &  $18.99$   \\
%\hline
 Ob(BP1) &  $0.9487$ &  $0.02518$ & $2.2 \times 10^{-9}$   &  $46.58$  &   $16.55$   \\
 \hline
 Ob(BP2) &  $0.953$ &  $0.0267$ & $2.2 \times 10^{-9}$   &  $59.4$  &   $30.77$   \\
 \hline
 Ob(BP3) &   \begin{tabular}{c}  $0.954$  \\\  $ 0.96$ \end{tabular} &   \begin{tabular}{c}  $ 0.0758$  \\\  $0.064$ \end{tabular} & $2.2 \times 10^{-9}$   &  \begin{tabular}{c}  $ 58$  \\\  $ 64$ \end{tabular}  &    \begin{tabular}{c}  $ 60$  \\\  $ 65.14$ \end{tabular}  \\
 \hline
  \hline
\end{tabular}
\end{center}
\caption{\label{tab:CMB-obs} \it CMB observables corresponding to BPs in Table~\ref{tab:BPs}}
\end{table}
Now we are ready to present the numerical results for the perturbation of the fields and  the power spectrum.  We choose three benchmark points (BPs) with a non-zero values of the inflaton vev, as shown in table~\ref{tab:BPs}. We have chosen consistent inflaton vev value that generates a reasonable peak in the power spectrum and induces GUT symmetry breaking at the scale $M_{\rm GUT} \geq 5\times 10^{15}$. We have calculated the CMB observables at the pivot scale $k_*= 0.05 \, {\rm Mpc}^{-1}$ as presented in Table~\ref{tab:CMB-obs}. We  solve the background equations numerically with inflation ending at $\epsilon_H=1$. For the initial velocity, we set $x'(0)=0$, and we chose the initial field value $x(0)$ such that the peak in the power spectrum occurs at $k\sim 10^{14}\, {\rm Mpc}^{-1}$, such that the PBH mass is larger than $10^{16}$ g. This fixes the inflation observable values shown in Table~\ref{tab:CMB-obs}.
\begin{figure}[H]
    \centering
    \includegraphics[width=0.8\linewidth]{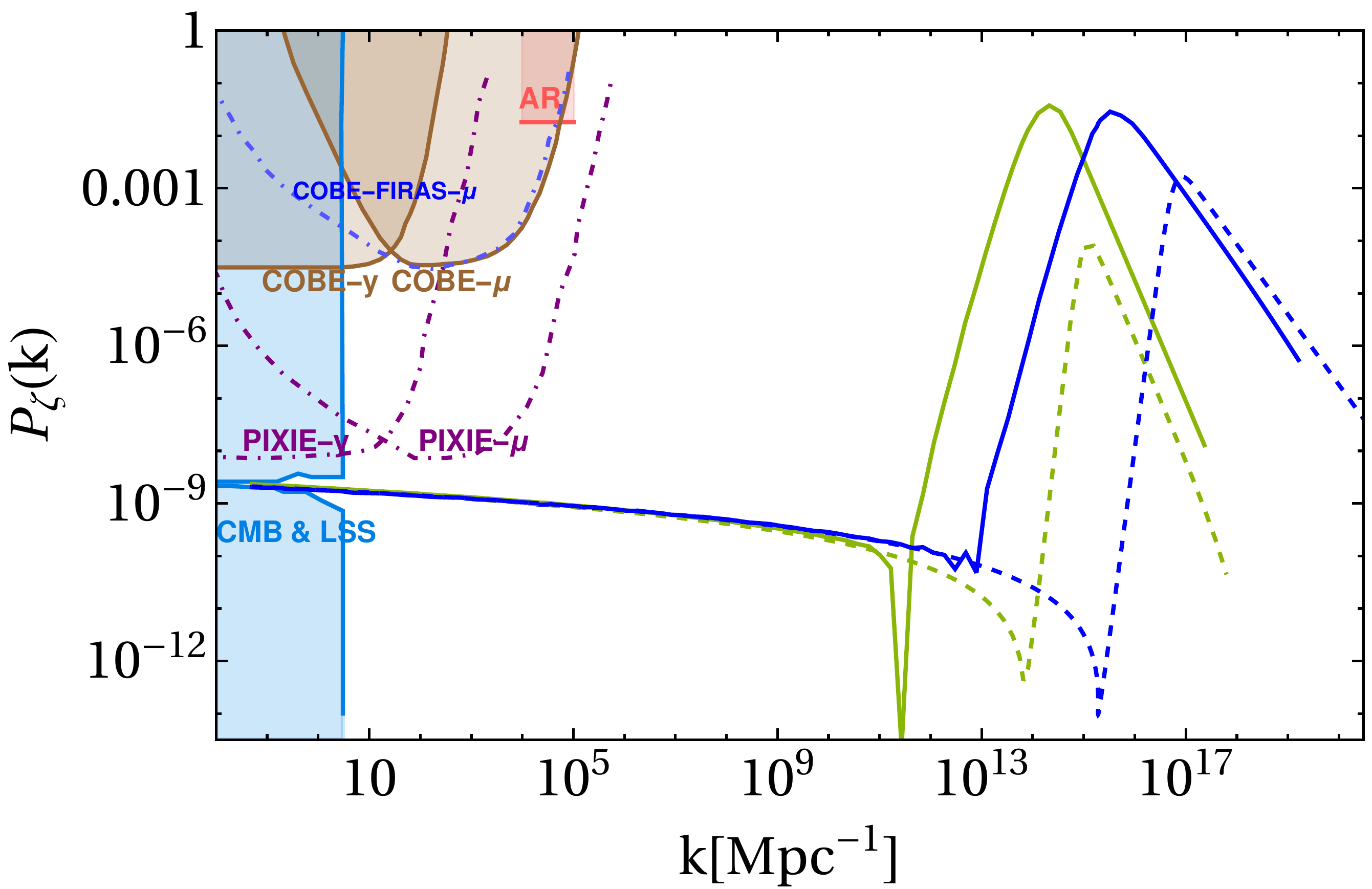}
\caption{\label{fig:ps1} \it The power spectrum calculated using the benchmark points in Table~\ref{tab:BPs}, solid green for $v=0.1 \, {\rm M_{Pl}}$, and solid blue for $v=0.8 \, {\rm M_{Pl}}$, by solving the exact equations for perturbations (Mukhanov-Sasaki Equations). The dashed curves represent the corresponding power spectra calculated in the slow-roll approximation. The shaded regions represent the constraints from current observations, while the dot-dashed curves represent future constraints ~\cite{Fixsen:1996nj,Chluba:2012we,Chluba:2013dna,Kogut:2011xw}.}
\end{figure}
In Fig.~\ref{fig:ps1}, we display the power spectrum $P_\zeta$ versus the scale $k$. As indicated in the figure, the curvature power spectrum is constrained at scales $10^{-4}\,\Mpc^{-1} \lsim k \lsim 1\,\Mpc^{-1}$, due to CMB observations.
We use two values of the inflaton vev, $v=0.1 \, {\rm and}\, 0.8 \, M_{\rm Pl}$, as well as the model parameters in Table~\ref{tab:BPs}. The solid curves are generated by numerically solving the MS equation~\ref{eq:remuk}, and the dashed curves are due to 
SR considerations.

\medskip
%
%%%%%%%%%%%%%%%%%%%%%%%%%%%%%%%%%%%%%%%%%%%%%%%%%%%%%%%%%%%%%%%%%
%\section{Scalar-induced Gravitational Waves as Probe of GUT scale}
\section{Scalar-induced Gravitational Waves and primordial black holes}
\label{sec:SIGW}
In this section, we study the scalar-induced gravitational waves (GWs) due to the enhancement in the primordial power spectrum. Indeed, second order tensor perturbations are sourced from enhanced scalar perturbations. The height of the power spectrum peaks as well as their position at scale, $k_{\rm peak}$, are dictated by the scalar potential parameters, including the inflaton vev $v$, which in turn is intimately related to the gauge symmetry breaking scale $M_{\rm SB}$, including the GUT scale. Therefore, these symmetry breaking scales can be probed via scalar-induced gravitational waves. Here, we assume that the gravitational waves were formed during the radiation-dominated epoch.

We follow \cite{Lewicki:2021xku,Kohri:2018awv,Espinosa:2018eve,Inomata:2019yww,Chatterjee:2017hru} in our numerical calculations of the spectrum of the scalar induced GWs. Using the primordial power spectrum  ${\cal P}_\zeta$, the the scalar induced GWs spectrum is given by 
%\begin{widetext}
\be \label{Omega_GW}
\Omega_{\rm GW}^{\rm si}h^2 \approx 4.6\times 10^{-4} \left(\frac{g_{*,s}^{4}g_{*}^{-3}}{100}\right)^{\!-\frac13} \!\int_{-1}^1 {\rm d} x \int_1^\infty {\rm d} y \, \mathcal{P}_\zeta\left(\frac{y-x}{2}k\right) \mathcal{P}_\zeta\left(\frac{x+y}{2}k\right) F(x,y) \bigg|_{k = 2\pi f} \,,
\ee
where
\bea
F(x,y) &=&\frac{(x^2\!+\!y^2\!-\!6)^2(x^2-1)^2(y^2-1)^2}{(x-y)^8(x+y)^8} \times \nonumber \\
&&
\!\left\{\left[x^2-y^2+\frac{x^2\!+\!y^2\!-\!6}{2}\ln\left|\frac{y^2-3}{x^2-3}\right|\right]^{\!2} \!+\! \frac{\pi^2(x^2\!+\!y^2\!-\!6)^2}{4}\theta(y-\sqrt{3}) \right\} .
\eea 
%\end{widetext}

%
\begin{figure}[H]
    \centering
    \includegraphics[width=0.8\linewidth]{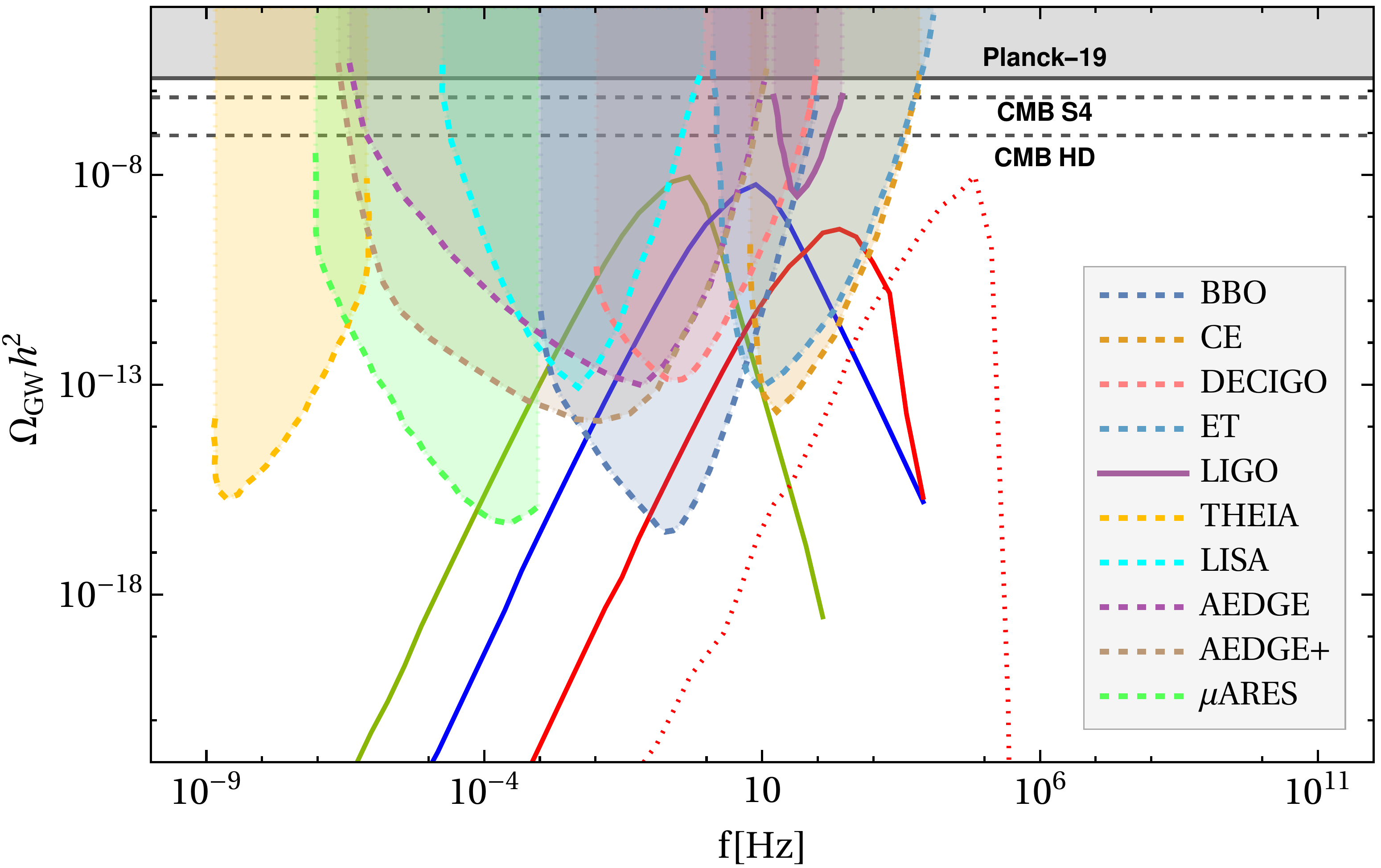}
    \caption{\label{fig:SIGW1}\it The  spectrum of scalar induced gravitational waves (solid red, blue and green curves for $v=0.1 , \, 0.8, \, 10 \, M_{\rm Pl}$) versus current and future GW detectors constraints. We have used the parameter values in Table~\ref{tab:BPs}. The red dotted curve corresponds to $v= 10 \, M_{\rm Pl}$, but with a modified value of the tensor to scalar ratio, $r\sim 0.064$. }
\end{figure}

After calculating the power spectrum as explained in the previous section, we feed it into Eq.~(\ref{Omega_GW}) in order to calculate the GWs energy density. In Fig.~\ref{fig:SIGW1} we plot the energy density of GWs, using Eq.~(\ref{Omega_GW}), versus the frequency.  The figure shows that the predicted GW spectra for the  parameter choices in Table~\ref{tab:BPs}, lie well within the detection range of future GW  experiments like LISA, DECIGO, BBO, SKA and ET~\cite{Audley:2017drz,Sato:2017dkf,Sathyaprakash:2009xs,Zhao:2013bba,Yagi:2011wg}. 
%Interestingly enough, we note that the recently reported NANOGrav \cite{Arzoumanian:2018saf,Aggarwal:2018mgp,Arzoumanian:2020vkk} signal can be interpreted in the context of this model  (purple lines). In this figure we display the NANOGrav 12.5 yrs  region.

  % 

%_________________________________________________________________________
%_________________________________________________________________________
%%%%%%%%%%%%%%%%%%%%%%%%%%%%%%%%%%%%%%%%%%%%%%%%%%%%%%%%%%%%%%%%%

%\section{Primordial Black Holes Abundance as Dark Matter}
%\label{sec:PBH}
Beside probing the GUT scale or other gauge symmetry breaking scales via the observable energy density of GWs, as described above, the significant enhancement of the scalar power spectrum can lead to copious production of PBHs. As shown in Ref.~\cite{DeLuca:2020agl}, the GWs spectrum  can be related to a prediction for the PBH abundance as DM.  Therefore, we calculate the mass of PBHs and their fractional energy density abundances. We will assume that the PBHs were formed during the radiation-dominated epoch, just like the gravitational waves. %\ag{Cite other dominated epoch papers..}

Following Ref. \cite{Spanos:2021hpk}, the fractional abundance of PBHs, $\Omega_ {PBH} / \Omega_ {DM}$, is defined to be:
\begin{equation}
\label{44}
\frac{\Omega_ {PBH}}{\Omega_ {DM}}(M_{\mathrm{PBH}})= 
\frac{\beta(M_{\mathrm{PBH}})}{8 \times 10^{-16}} \left(\frac{\gamma}{0.2}\right)^{3/2} 
       \left(\frac{g_*(T_f)}{106.75}\right)^{-1/4}\left(\frac{M_{\mathrm{PBH}} \,  }{10^{-18 }\; \mathrm{ grams} }\right)^{-1/2}\, , 
\end{equation}
where $M_{\rm PBH}$ denotes the PBH mass,  $\Omega_{DM} \simeq 0.26$ is the  DM abundance, and the factor $\gamma$ represents the dependence on the gravitation collapse and is set to be equal to $0.2$~\cite{Carr:1975qj}. The function $\beta(M_{\rm PBH})$ shows the mass fraction of Universe collapsing into PBH. $T_f$  represents the temperature  at which PBHs are formed, and  $g_*(T_f)$ denotes the effective degrees of freedom during the formation of PBHs. The fractional abundance of PBHs $f_ {PBH}$ is then given by \cite{Spanos:2021hpk}
\be
f_ {PBH} = \int \frac{d  M_{\mathrm{PBH}}}{M_{\mathrm{PBH}}} \, \frac{\Omega_ {PBH} }{\Omega_{DM}}\,  .
\label{fpbh}
\ee
After inflation ends, the modes re-enter the Hubble horizon $H^{-1}$ and PBHs are formed. With the assumption of  spherical collapse of perturbations, we have ~\cite{Ballesteros:2017fsr}

\be
M_{\mathrm{PBH}}=\gamma \frac{4\, \pi \,  \rho}{3}  H^{-3} \,. 
\ee
Here $\rho$ is the energy density of Universe during collapse to form PBHs. For PBHs created during the radiation epoch, the PBHs mass is given (in grams) as a function of the co-moving wavenumber $k$ \cite{Ballesteros:2017fsr,Spanos:2021hpk}, as follows  
%~\cite{Spanos:2021hpk}
%\begin{equation}
%M_{\mathrm{PBH}}=\gamma \frac{4\, \pi \, \rho}{3}\,  H_{m-r}^{-3} \left( \frac{g(T_f)}{g(T_{m-r})}\right)^{1/2}\left( \frac{g_s(T_f)}{g_s(T_{m-r})}\right)^{-2/3}\left( \frac{k}{k_{m-r}}\right)^{-2}  \, .
%\end{equation}
%The subscript ${m-r}$ denotes  the time  of matter and radiation equality, and $g_s$ refers to the entropy density. 
%The last equation is a result of the mere entropy  conservation $d(g_s(T)T^3a^3)/dt=0$ between the epoch of the re-entry of the co-moving wavenumbers and the epoch when radiation-matter equality takes place. 
%
\begin{equation}
\label{43}
M_{\mathrm{PBH}}(k)=10^{18} \left( \frac{\gamma}{0.2} \right) \left(\frac{g_*(T_f)}{106.75}\right)^{-1/6} \left(\frac{k}{7 \times 10^{13} \, \mathrm{Mpc}^{-1}  }\right)^{-2}  .
\end{equation}
%
%where we have used the common approximation    $g(T)=g_s(T)$ ~\cite{Ballesteros:2017fsr}. 
The factor ${g_*(T)=106.75}$ in case we assume the SM spectrum.  However,  with a spectrum of SU(5) GUT as we have in our scenario, we set  $g_*(T)=228.75$. Thus the PBH fractional abundance in the SM is 1.13 times larger than in the SU(5). This relative factor, to a good approximation, can be safely ignored. 
  
%\item
%
\begin{figure}[H]
    \centering
    \includegraphics[width=0.8\linewidth]{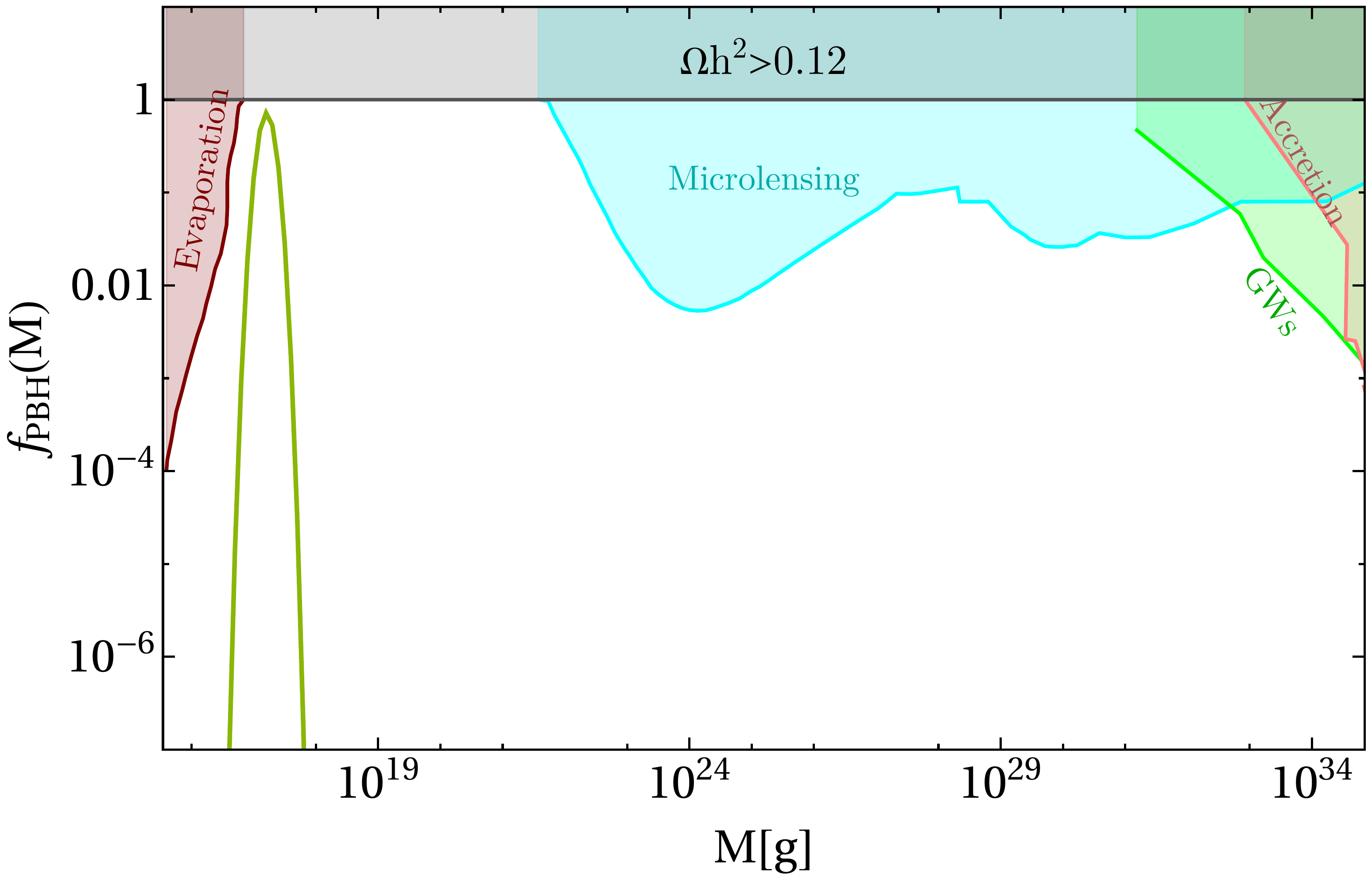}
    \caption{\label{fig:PBH1}\it The fractional abundance of PBHs as a function of their mass corresponding to the parameter values in Table~\ref{tab:BPs}, with the inflaton vev  $v= 0.1\, M_{\rm Pl} $.  The shaded regions correspond to the observational constraints \cite{Carr:2009jm,Green:2020jor,Carr:2020gox}. PBHs generated by other benchmark points with larger vev values, such as $v= 0.8\, M_{\rm Pl} $ and $v= 10\, M_{\rm Pl} $, are expected to evaporate before BBN.
    }
    
\end{figure}
 We follow  the Press-Schechter approach in order to evaluate the mass fraction $\beta$. Assuming that, the overdensity $\delta$ follows a gaussian probability distribution function, $\beta$ can be calculated from the following integral
\begin{equation}
\label{42}
\beta(M_{\mathrm{PBH}})= \frac{1}{\sqrt{2 \pi \sigma ^2 (M_{\mathrm{PBH}})}} \int^{\infty}_{\delta_c} d\delta \,  \exp \left(  -\frac{\delta ^2}{2 \sigma^2(M_{\mathrm{PBH}}) } \right) \, , 
\end{equation}
with $\delta_c$ being a threshold for the PBH collapse, and
 $\sigma$ is the variance of the curvature perturbation, which can be written in terms of the co-moving wavenumber as follows
\begin{equation}
\label{40}
\sigma^2 \left( M_{\mathrm{PBH}}(k)  \right)= \frac{16}{81}  \int \frac{dk' }{k'} \left(\frac{k'}{k}\right)^4 P_{R}(k') \widetilde{W}\left(\frac{k'}{k}\right),
\end{equation}
 where the window function  $\widetilde{W}(x)$ can be approximated with a Gaussian distribution function to be: $ \widetilde{W}(x)=e^{-x^2/2} $. For $\delta_c$, following Refs.~\cite{Harada:2013epa,Musco:2008hv,Musco:2004ak,Musco:2012au,Musco:2018rwt,Escriva:2019phb,Escriva:2020tak,Musco:2020jjb}, we have taken its values in the range between 0.4 and 0.6.

Figure~\ref{fig:PBH1} depicts the fractional abundance of the primordial black holes as a function of their mass. We have used only one benchmark point (BP) for the parameters of Table~\ref{tab:BPs}, corresponding to $v=0.1 M_{\rm Pl}$. The PBHs generated by other benchmark points with larger vev values, such as $v= 0.8\, M_{\rm Pl} $ and $v= 10\, M_{\rm Pl} $, are expected to evaporate before BBN, and may not serve as the DM candidate (whole or partially).
The figure shows that PBH abundance can account for the observed DM relic density for PBH mass scale $\sim 10^{17} \, {\rm g} \simeq 7.5 \times 10^{-17} \, M_\odot$, where $ M_\odot$ is the solar mass. The shaded regions are disallowed by observations from Planck, Accretion disk, Microlensing, Gravitational Waves and black hole evaporation and several other constraints \cite{Carr:2009jm,Green:2020jor,Carr:2020gox,Inoue:2017csr,Montero-Camacho:2019jte,Katz:2018zrn,Poulin:2017bwe,Capela:2013yf,Niikura:2017zjd,Wyrzykowski:2011tr,Griest:2013esa,Tisserand:2006zx,Ali-Haimoud:2016mbv,Gaggero:2016dpq}. 

%\ag{We need to write more details regarding the observations constraints.}

\section{Discussion and Conclusions}
In this paper we present a single-field inflection-point model of SU(5) inflation. The inflaton is an SU(5) gauge-singlet which couples to the SU(5) Higgs responsible for SU(5)-symmetry breaking induced via the Higgs-portal mixed quartic coupling. As the inflaton rolls down to its potential minimum, the SU(5) symmetry is also broken leading to a one-to-one correspondence between the GUT scale and the scale of inflection point via the inflaton vev. We show that such an inflationary scenario may lead to the generation of both detectable GWs and sufficient PBHs as the sole DM candidate of the universe. We summarize our main findings below:
\begin{itemize}
    \item The scale of grand unification which is otherwise very challenging to probe in laboratory experiments is highly constrained. We investigated a cosmological scenario in SU(5) which provides a pathway to probe the SU(5) gauge symmetry breaking scale via inflationary observables. We show this for a  model involving inflection-point inflation employing an SU(5) singlet Higgs non-minimally coupled to the Ricci scalar. We predict the CMB observables with respect to the choice of GUT scale $M_{\rm GUT}$ (see Table \ref{tab:CMB-obs}). The  prediction r $\approx$ 0.026 for  \textit{tensor-to-scalar} ratio, will be within reach of the next generation CMB experiments for our choice of the benchmark point.
    \item Since the SU(5) scalar singlet acquires a vev and mixes with the SU(5) adjoint higgs which breaks symmetry to the SM, we expect quantum corrections to the self-quartic couplings from gauge interactions. Constraints on such couplings from quantum corrections in this scenario have been estimated (see section \ref{sec:pt-infl}). We also discuss implications for models based on gauge groups such as $SU(4)_C \times SU(2)_L\times SU(2)_R$ and  $SU(3)_C \times SU(3)_L\times SU(3)_R$, which can accommodate symmetry breaking scales lower than $M_{\rm GUT}$.
    \item We estimate the power spectrum across all k-values which provides constraints on the SU(5) model from the measurements of CMB spectral distortions (see Fig. \ref{fig:ps1}). The power spectrum shows a spike in the amplitude which corresponds to the SU(5) symmetry breaking scale via its relation with the inflection-point and the inflaton rolling to its minima and completing the SU(5) symmetry breaking.
    \item Second-order tensor perturbations propagate as GW that are detectable with $\Omega_{\rm GW}h^2 \sim 10^{-9}$ and peak frequency f $\sim$ 0.1 Hz by LISA and $\Omega_{\rm GW}h^2 \sim 10^{-10}$ and peak frequency of $\sim$ 10 Hz in ET. Furthermore, in other next generation GW observatories such as AEDGE, BBO, DECIGO, one may be able to detect this signal, and this will act as a novel probe of GUT scale physics (see Fig. \ref{fig:SIGW1}). 
    \item Production of PBH of masses $10^{17}-10^{18}$g ($ 10 - 100 \Msun$) as the sole DM candidate in the Universe is proposed. This novel DM candidate is also a signature of scale of grand unification involving inflationary cosmology (see Fig. \ref{fig:PBH1}). 
\end{itemize}
We believe that the precision that GW astronomy aspires from the planned worldwide network of GW detectors can make the dream of testing high-scale and fundamental BSM scenarios of UV-completion like Grand Unification a reality which complement the laboratory searches in near future.

\medskip

\section*{Acknowledgments}
The work of AM is  supported in part by the Science, Technology and Innovation Funding Authority (STDF) under grant No. 33495. The authors thank Amit Tiwari and Shiladitya Porey for careful reading of the manuscript.

%\newpage

%\medskip

\end{document}